\documentclass[journal]{IEEEtran}
\usepackage{cite}
\ifCLASSINFOpdf 
\usepackage[pdftex]{graphicx}
\else
\usepackage[dvips]{graphicx}
\fi

\usepackage[cmex10]{amsmath}
\usepackage{algorithmic}
\usepackage[ruled,norelsize]{algorithm2e}

\usepackage{array}
\usepackage{mdwmath}
\usepackage{mdwtab}
\usepackage{todonotes}

\usepackage[vertfit]{breakurl} 
\usepackage{eqparbox}
\ifCLASSOPTIONcompsoc
\usepackage[caption=false,font=normalsize,labelfont=sf,textfont=sf]{subfig}
\else
\usepackage[caption=false,font=footnotesize]{subfig}
\fi
\usepackage{fixltx2e}
\usepackage{url}
\usepackage{amsfonts}
\usepackage{amssymb}
\usepackage{amsthm}
\usepackage{bm}
\usepackage[ruled]{algorithm2e}
\usepackage[utf8]{inputenc}
\usepackage{booktabs}
\usepackage{url}
\usepackage{cite}
\usepackage{epstopdf}

\newcounter{tempEquationCounter} 
\newcounter{thisEquationNumber}

\begin{document}
\title{On Performance Evaluation of \\Random Access Enhancements for 5G uRLLC}


%\author{Adnan~Aijaz,~\IEEEmembership{Member,~IEEE}
%        %Shuyu~Ping,
%        %Adnan~Aijaz,~\IEEEmembership{Member,~IEEE,}
%        %and~A.~Hamid~Aghvami,~\IEEEmembership{Fellow,~IEEE}% <-this % stops a space
%\thanks{The author is with the Telecommunications Research Laboratories, Toshiba Research Europe Ltd., Bristol, BS1 4ND, UK. Contact e-mail: adnan.aijaz@toshiba-trel.com}}

%\author{\IEEEauthorblockN{Adnan Aijaz}\\
%\IEEEauthorblockA{Telecommunications Research Laboratories,  Toshiba Research Europe Ltd., Bristol,  UK.\\
%adnan.aijaz@toshiba-trel.com}}

\author{Jayashree Thota and Adnan Aijaz \\
Telecommunications Research Laboratory,
Toshiba Research Europe Ltd., Bristol, BS1 4ND, U.K. \\
\(\left\lbrace \textrm{jaya.thota, adnan.aijaz} \right\rbrace\)@toshiba-trel.com\\ % <-this % stops a space
\thanks{}%
}
\maketitle
\begin{abstract}
\boldmath
%One of the key challenges to successfully deliver ultra-reliable low-latency communications (uRLLC) is the need to enhance the operation of the cellular random access channel (RACH). The state-of-the-art LTE RACH procedure is challenging due to the severe congestion and overloading from massive machine type communication (mMTC) devices leading to collisions especially in a densely populated factory scenario. This paper proposes novel RACH enhancements for uRLLC devices to significantly lower the collision probability resulting in shorter access delay and improved reliability, compared with the standard LTE random access (RA) scheme. Early data transmissions, parallel preambles, dynamic reserved preambles, enhanced back-off, 5G flexible numerology are proposed to satisfy the stringent requirements of uRLLC devices. Performance evaluation, based on comprehensive validated system-level simulator, demonstrates that the proposed enhancements can fulfil the 3GPP uRLLC target RA access delay of less than 10ms with 99.99\% reliability  in factory automation environments. 
One of the key challenges in realizing ultra-reliable low-latency communications (uRLLC) for factories-of-the-future (FoF) applications is to enhance the cellular random access channel (RACH) procedure. The state-of-the-art LTE RACH procedure does not fulfil the latency requirements for envisioned FoF applications. Moreover, it becomes challenging due to  congestion and overloading from massive machine type communication (mMTC) devices leading to collisions especially in a densely populated factory scenarios. The main objective of this paper is to conduct a comprehensive performance evaluation of different random access (RA) enhancements for uRLLC over 5G wireless networks. Our performance evaluation is based on a realistic system-level simulator. The core enhancements considered in this work include early data transmission (EDT), reserved preambles and the use of flexible physical (PHY) layer numerology. We also propose three new RA enhancements for uRLLC. Performance evaluation demonstrates that the proposed RA enhancements can fulfil the 3GPP control plane target  of less than 10 ms latency with 99.99\% reliability  in factory environments. 
%This paper proposes novel RACH enhancements for uRLLC devices to significantly lower the collision probability resulting in shorter access delay and improved reliability, compared with the standard LTE random access (RA) scheme. Early data transmissions, parallel preambles, dynamic reserved preambles, enhanced back-off, 5G flexible numerology are proposed to satisfy the stringent requirements of uRLLC devices. Performance evaluation, based on comprehensive validated system-level simulator, demonstrates that the proposed enhancements can fulfil the 3GPP uRLLC target RA access delay of less than 10ms with 99.99\% reliability  in factory automation environments. 
%\vspace{9pt}
%\emph{Index Terms}---Cognitive Radio Ad-Hoc Networks, Cooperative Routing, Spectrum %Aggregation

\end{abstract}



% \fi
\IEEEpeerreviewmaketitle

\section{Introduction}
\IEEEPARstart{T}{he} emerging 5G wireless networks are expected to support diverse use-cases which can be broadly classified into three categories \cite{ITU_2083} enhanced mobile broadband (eMBB), massive machine type communications (mMTC) and ultra-reliable low-latency communications (uRLLC). Owing to stringent reliability and latency targets, the most challenging design requirements are created by uRLLC which is the key enabler for the various critical applications across different vertical industries \cite{TI_JSAC}.

The recent Industry 4.0 initiative aims at enhancing the versatility, flexibility and  productivity of legacy industrial systems to create highly efficient, connected, flexible and self-organized factories, often referred to as factories-of-the-future (FoF) in the manufacturing sector. A robust and ubiquitous connectivity layer supporting uRLLC is essential in realizing the FoF vision.  Typical uRLLC applications in FoF include motion control for moving or rotating parts of machinery, collaborative operation of mobile robots, mobile control panels with safety function, time-critical process optimization to support zero-defect manufacturing, real-time monitoring, and remote maintenance \cite{haerick20155g}. The third generation partnership project (3GPP) aims at realizing such uRLLC applications with ultra-low latency of 1 ms and 10 ms for user plane and control plane, respectively, and ultra-high reliability of more than 99.999\% in terms of packet delivery performance \cite{krause2016study}. However, the most critical source of latency in state-of-the-art long term evolution (LTE) radio access networks (RAN) is the initial link establishment using random access channel (RACH) procedure that can take several tens of milliseconds \cite{chen2018ultra}. This becomes particularly problematic for FoF applications due to intermittent transmissions by uRLLC devices with small payloads contending for fixed number of preambles along with the massive, periodic and bursty nature of other applications. This leads to severe congestion at the LTE medium access control (MAC) layer, especially in dense factory environments.

\subsection {State-of-the-Art}
In literature, several techniques have been proposed to reduce the random access (RA) delay. 3GPP has proposed early data transmission (EDT) as part of the Release 15 specification. EDT is one of the most attractive technique to reduce the connection setup signaling overhead and shorten the overall transmission time. In EDT, the uplink grant for data is sent early, thus allowing data transmission to be piggybacked with RACH procedure. Hoglund \emph{et al.} \cite{hoglund20183gpp} provided some initial results on EDT performance which show that it exhibits gains in terms of battery life improvement by up to 46\% and reduction of message latency by 85 ms at the cell edge. Performance studies conducted by Condoluci \emph{et al.} \cite{condoluci2016enhanced} show that a RACH procedure based on two-message handshake through a specially designed preamble set  can guarantee a delay reduction from 10\%-50\% (in case of a macro cell) and 50\%-70\% (in case of a femto cell) depending on the load as compared to the standard RACH procedure. 

Chen \emph{et al.} \cite{chen2017prioritized} proposed separate RACH resources for uRLLC and eMBB traffic which is termed as prioritized resource reservation. Simulation results show an access delay below 10 ms for 95\% of uRLLC devices can be obtained by reserving preambles at least double the number of incoming uRLLC requests. 
%However, in the case of a factory setup with thousands of devices, reserved preambles can waste the valuable RACH resources and increase collisions of non-uRLLC devices. 

Results from system level simulations \cite{ashraf2016ultra} show that LTE wireless systems cannot support stringent latency requirements of uRLLC applications and 5G new radio (NR) with flexible physical (PHY) layer numerology is essential. Access class barring (ACB) with different back-offs depending upon the traffic priority has been introduced in LTE \cite{3GPP37868}. Simulations results using an analytical model in  \cite{leyva2017accurate} show that ACB does not satisfy the 3GPP control plane requirements for uRLLC. 

Diversity in the form of repeated transmissions for contention-based RA was proposed by 3GPP for narrow band internet of things (IoT) devices to improve the reliability. The authors in \cite{jiang2018rach} developed a stochastic geometry framework to analyse effect of diversity by repeating the preamble to improve the RACH success probability. Results show that preamble repetitions can result in inefficient channel resource utilization in a heavy traffic scenario. Also, Vural \emph{et al.} \cite{vural2017success} show that benefits of using multiple preamble RACH procedure can be seen for lower preamble set size (up to 20) as the channel saturates with repeated transmissions.

In summary, RA enhancements to reduce the access delay include introduction of short transmission slots via 5G NR numerology, allowing faster uplink data transmissions by EDT, lower back-off timers for high priority devices, and reserving resources for uRLLC applications. However, these are the potential candidates and have not been validated in FoF scenarios. Also, none of these techniques ensure successful RA in a single attempt. Ensuring reliability requires more radio resources (e.g., parity, redundancy via diversity, and re-transmissions), albeit increasing latency over sub-millisecond (ms) target for the uRLLC applications. In this paper we propose novel RA enhancements including parallel preambles, dynamic reserved preambles and enhanced back-off to reduce the collision probability. Simulation results show that our proposed techniques combined with EDT not only satisfy the 3GPP stringent latency requirements but also guarantee the reliability targets for uRLLC applications in the control plane for FoF applications.

\subsection {Contributions and Outline}
To this end, this paper has a two-fold objective. Firstly, it conducts a comprehensive performance evaluation of different RA enhancements for uRLLC. Secondly, it develops new RA enhancements for FoF-centric uRLLC applications. The main contributions of this work are summarized as follows.

\begin{itemize}
\item We provide an overview of the existing RA enhancements that are particularly attractive for uRLLC applications. Such enhancements include EDT, the use of flexible numerology and reserved preambles. 

\item We propose three new RA enhancements for uRLLC which have been designed to fulfil the  requirements of FoF applications. The proposed enhancement include parallel preamble transmission, enhanced back-off and dynamic reserved preamble techniques. 

\item We develop a realistic system-level simulator to evaluate different RA enhancements. The simulator has been validated against the widely used 3GPP model \cite{3GPP37868}.

\item We conduct a comprehensive performance evaluation of the existing as well as proposed RA enhancements. Performance has been bench-marked against standard LTE RA procedure. 

\end{itemize}

The remainder of this paper is structured as follows. In Section II, the state of the art RACH procedure is discussed. A detailed description of existing and proposed RACH enhancements are presented in section III and IV respectively. The system model and results are discussed in sections V and VI with a conclusion in section VII.      
%%%%%%%%%%%%%%%%%%%%%%%%%%%%%%%%%%%%%%%%%%%%%%%%%%%%%%%%%%%%%%%%%%%%%%%%%%%%%%%%%%%
\section{Standard RACH Procedure}
\begin{figure}
   \centering
        \includegraphics[scale=0.45]{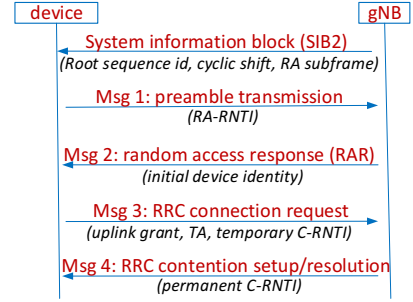}
      \caption{Standard RACH procedure.}
      \label{std_LTE}
\end{figure}

 The state-of-the-art LTE RACH procedure \cite{3GPP3600} is shown in Fig.  \ref{std_LTE}. In system information block (SIB2), the next generation node B (gNB) periodically broadcasts several parameters such as root sequence ID, RACH configuration index, power offset, and initial power. In a contention based RACH procedure, the device randomly selects a preamble out of the 54 orthogonal zadoff-chu (ZC) sequences generated by root sequence cyclic shift. This is transmitted as Msg 1 on the RA subframe in time and resource block (RB) in frequency implicitly defining the RA-radio network temporary identifier (RA-RNTI). The gNB responds with Msg 2 random access response (RAR) containing a temporary cell-RNTI (C-RNTI), timing advance (TA) and uplink resource grant upon Msg 1 success. In Msg 3, the device transmits a radio resource control (RRC) connection request including a randomly chosen initial device identity after decoding the RB assignment from Msg 2. Multiple devices can select the same preamble, RA-RNTI in Msg 1 and also the corresponding C-RNTI in Msg 2 and transmit their own Msg 3 on the uplink resources which is detected as a collision by gNB. In Msg 4, the gNB sends RRC connection setup with a permanent C-RNTI and an echo of the initial identity transmitted in Msg 3 by the device. RACH procedure is considered as a success if the identities are matched else the device retries the procedure after a back-off interval. The successful device is ready to transmit uplink data. 
 
%%%%%%%%%%%%%%%%%%%%%%%%%%%%%%%%%%%%%%%%%%%%%%%%%%%%%%%%%%%%%%%%%%%%%%%%%%%%%%%%%%
\section {Existing RA Enhancements for uRLLC}
The core enhancements to reduce the access delay include EDT and 5G NR flexible numerology. Reserved preambles are proposed to reduce the Msg 1 collision probability for uRLLC devices in case of a mixed traffic scenario. 
\subsection {EDT}
\begin{figure}
   \centering
        \includegraphics[scale=0.45]{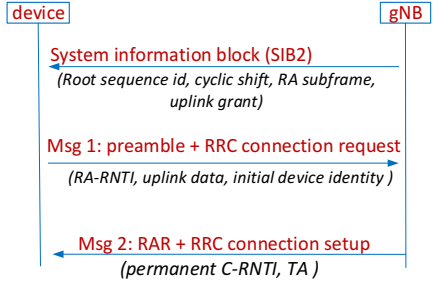}
      \caption{Two-step EDT procedure. }
      \label{EDT}
\end{figure}

EDT was proposed by 3GPP in \cite{EDT} for uRLLC devices to lower the access delay in the control plane. It is generally a two-step RACH procedure where, EDT Msg 1 carries the standard LTE RACH Msg 1 and Msg 3 i.e. preamble followed by the data (connection request, device ID, buffer status report) as shown in Fig.  \ref{EDT}. EDT Msg 2 corresponds to the Msg 2 and Msg 4 of standard LTE RACH i.e. the RAR, TA and finally the connection complete with RRC response message. It is assumed that the resource allocation for data in EDT Msg 1 uses uplink shared channel that is pre configured by the gNB in the SIB2 prior to the start of the EDT process. Also, the transmission of the information part is done right after the guard time (GT) period of the preamble which acts as a TA window to ensure preamble reception. 
   
\subsection{Reserved Preambles}
\begin{figure}[!htbp]
   \centering
        \includegraphics[clip,scale=0.6]{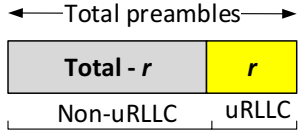}
      \caption{Illustration of the preamble reservation technique.}
      \label{res}
   \end{figure}

Preambles can be reserved for uRLLC devices as suggested in \cite{chen2017prioritized}. Fig.  \ref{res} shows the division of contention based preambles (a total of 54) between uRLLC devices and non-uRLLC devices. Priority is set via preamble reservation for uRLLC devices and the total number of reserved preambles is given by \(r\). Reserving preambles in case of a mixed traffic scenario can reduce collision probability of uRLLC devices. However, as uRLLC devices transmit intermittently, unused reserved preambles can waste the valuable resources and increase the collision probability of non-uRLLC devices.  

\subsection{Flexible Numerology}
3GPP has introduced a scalable and flexible frame structure for 5G NR \cite{pocovi2018achieving} which can shorten the transmission time interval (TTI) duration as compared to LTE. The subcarrier spacing (15 kHz for LTE) is configurable to 30/60/120 kHz in the frequency domain. The number of symbols per slot (7 symbols for LTE) can also be configured to mini-slots with 4 or 2 symbols. Such flexible numerology has potential to significantly lower the access delay.
%%%%%%%%%%%%%%%%%%%%%%%%%%%%%%%%%%%%%%%%%%%%%%%%%%%%%%%%%%%%%%%%%%%%%%%%%%%%%%%%%%%
\section {Proposed RA Enhancements for uRLLC}
EDT is proposed to reduce the RA access delay; however, it does not reduce the Msg 1 collision probability. Thus, using EDT as the core enhancement, we propose novel RA enhancements to reduce the Msg 1 collision probability and ensure successful RA in a single attempt.

\subsection {Parallel Preamble Transmissions}
\begin{figure}
   \centering
        \includegraphics[scale=0.5]{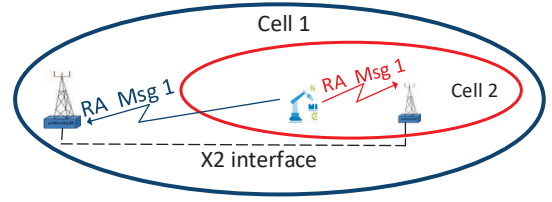}
      \caption{Parallel preamble using dual-connectivity.}
      \label{DC}
 \end{figure}

Dual Connectivity (DC) allows a user to be simultaneously served by two different base stations, operating on two different carrier frequencies, and connected via a non-ideal back-haul \cite{3GPP36842}. DC is generally applicable to a UE in connected mode. However, it can also be exploited in the idle mode. In idle mode, a DC-capable device can perform RACH procedure on both master-gNB (MgNB) and secondary-gNB (SgNB). This is termed as  parallel preamble transmission and illustrated in Algorithm 1, where \(P_{MgNB}\)  and \(P_{SgNB}\) are the preamble sets of the MgNB and the SgNB, respectively. The device randomly chooses a preamble from each of the sets and transmits Msg 1 independently and simultaneously on both the gNBs. In the factory environment, the two gNBs can differ in transmit power, as shown in Fig.  \ref{DC}, where Cell 1 provides higher coverage than Cell 2. It is assumed that device uses different RA-RNTIs in Msg 1 as it selects different preambles on both the gNBs. Uplink resources can be allocated by either gNB or both gNBs where, the device responds to the first RAR Msg 2 from either of them. This provides a significant reduction in message collision probability, as it is less likely for the device to pick the same preamble from different preamble sets.

 \begin{algorithm}
   \caption{Parallel preamble  transmissions.}
   \textbf{Input}: \(P_{MgNB},P_{SgNB}\) \\
   \textbf{Output}: \(RAR\) \\
    Randomly choose p1 from \(P_{MgNB}=[1,2 \ldots 54]\) \\
    Randomly choose p2 from \(P_{SgNB}=[1,2 \ldots 54]\) \\
    Transmit Msg 1 with p1 to \(MgNB\) and p2 to \(SgNB\)  \\
    Wait \(T_{Msg\hspace{1mm}2}\) = 3 ms to receive RAR  \\
    \uIf {RAR received from both gNBs}{
     Device processes first received RAR \\ }
    \uElseIf {RAR received from one gNB}{
     Device processes received RAR \\ }
    \Else {
     Device restarts RA procedure after back-off \\
    }
\end{algorithm}  

\subsection {Enhanced Back-off}
This enhancement includes reducing the default RA response window (\(RAR_{window}\)) and back-off indicator (\(BI\)) for all RACH failed devices as shown in Algorithm 2, where \(F_{uRLLC}\) and \(F_{non-uRLLC}\) represent the Msg 1 failed uRLLC and non-uRLLC devices respectively. The failed uRLLC devices have higher priority with \(BI\) down to 0 ms as compared to failed non-uRLLC devices with \(BI\) down to 10ms. The standard \(BI\) is 20 ms. The device detects its failure of sending Msg 1 after the processing delay time of \(T_{Msg\hspace{1mm}2}\) = 3 ms and the back-off timer (i.e. \(T_{Msg\hspace{1mm} 2} + RAR_{window} + BI\)) and will re-attempt the RACH procedure. This enhancements ensures that the RACH procedure is attempted early giving priority. 

\begin{algorithm}
   \caption{Enhanced back-off (EBF).}
   \textbf{Input}: \(F_{uRLLC},F_{non-uRLLC},EBF\) \\
   \textbf{Output}: \(BI,RAR_{window}\) \\
    \uIf {EBF ==`true' and \(F_{uRLLC}\) ==`true'}{
     \(RAR_{window}\) = 0, \(BI\) = 0 \\}
    \uElseIf {EBF ==`true' and \(F_{non-uRLLC}\) ==`true'}{
     \(RAR_{window}\) = 0, \(BI\) = 10 \\}
    \Else {
     \(RAR_{window} = 5\), \(BI = 20\) \\ 
    }
\end{algorithm}  

\subsection {Dynamic Reserved Preambles}
Usually, the number of reserved preambles is broadcast by the gNB periodically. In dynamic reserved preambles, instead of fixed reserved preambles, i.e., \(r\) = 3, the gNB updates the reserved preambles by calculating the moving average of the number of devices in the priority list in the prior SIB2 period (usually 80 ms) as shown in Algorithm 3, where \(N_{uRLLC},N_{non-uRLLC}\) represent the new uRLLC and non-uRLLC device interested in RACH, \(F_{uRLLC},F_{non-uRLLC}\) are the uRLLC and non-uRLLC devices that failed RACH previously. Normally, the reserved preambles are used by only uRLLC devices. However, in the proposed dynamic reserved preambles (DRP) enhancements, the variable set of reserved preambles are contended by priority devices \(K\) which include both the uRLLC and the previously Msg 1 failed non-uRLLC devices. This dynamic allocation between the reserved and non-reserved preambles reduces the collisions between the non-uRLLC devices.  
%%%%%%%%%%%%%check DRP enhanced backoff algo %%%%%%%%%%%%%%
 %\caption{My algorithm}
% \removelatexerror
  \begin{algorithm}
   \caption{Dynamic reserved preamble.}
   \textbf{Input}: \(N_{uRLLC},N_{non-uRLLC},F_{uRLLC},F_{non-uRLLC},DRP\) \\
   \textbf{Output}: \(r\) \\
    \For {each SIB2 period \(T_{SIB2}\)=80 ms} {
     \For {each RA subframe = 5 ms}{
        \uIf {DRP==`true'}{
          \(K\) = \{\(N_{uRLLC},F_{uRLLC},F_{non-uRLLC}\)\} \\
          \(r\) = moving average (\(K\)) \\ 
          }
        \Else {
          \(K\) =  \{\(N_{uRLLC},F_{uRLLC}\)\} \\
          \(r\) = 3 \\
        }
    }
    }
\end{algorithm}
%%%%%%%%%%%%%%%%%%%%%%%%%%%%%%%%%%%%%%%%%%%%%%%%%%%%%%%%%%%%%%%%%%%%%%%%%%%%%%%%%%%

\section{System Model}
\begin{figure}[!htbp]
   \centering
        \includegraphics[scale=0.45]{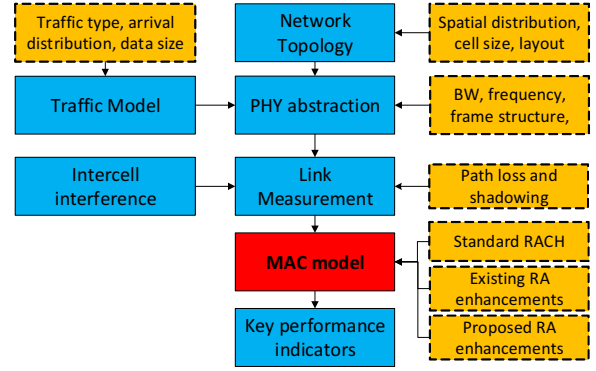}
      \caption{Schematic block diagram of the simulator.}
      \label{system}
\end{figure}

\begin{figure}[!htbp]
   \centering
        \includegraphics[scale=0.4]{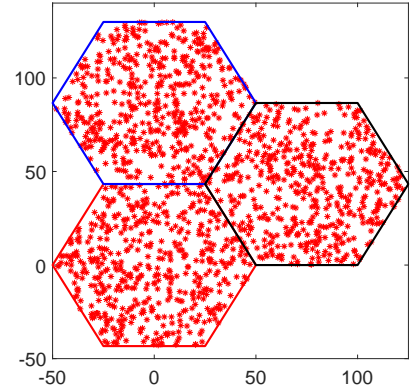}
      \caption{Cell layout.}
      \label{cell}
\end{figure}

The schematic block diagram of the simulator is shown in Fig.  \ref{system}. 
In the network topology block, devices are uniformly distributed in a three cell hexagonal layout with a maximum cell radius of 50 meters (m) as shown in Fig.  \ref{cell}. An indoor gNB at the centre of each cell is considered. Two types of devices, i.e., uRLLC and non-uRLLC devices are uniformly distributed within the cell. In the traffic model block, the total number of devices and their arrival distribution is given in Table \ref{parameters}. The  PHY abstraction block creates a PHY layer model based on channel bandwidth (BW), frequency, transmit power, frequency division duplex (FDD) type 1 radio frame structure as given in Table \ref{parameters}. We adopt an indoor propagation model from \cite{tanghe2008industrial}, where the path loss (PL) at a reference distance of 15m is 63.57 dB and a PL exponent of 3.44 is used. 
%Also, based on an indoor propagation model \cite{tanghe2008industrial}, the path loss (PL) at a reference distance of 15m (\(PL(d_{0})\)) = 63.57 dB and a PL exponent (\(n\)) = 3.44 is used in the path loss block. 
The device signal to interference plus noise ratio (SINR) depending upon the received power, the noise power and the interference is calculated for each device. The SINR of devices from adjacent cells is used as inter-cell interference block. 

The main block is the MAC model where the RACH procedure discussed in section II with existing enhancements in section III and proposed enhancements in section IV is evaluated. In case of a collision, where two (or more) devices select the same preamble at the same time, it is assumed that the gNB will not be able to decode any of the preambles; hence, the gNB will not send the Msg 2 RAR. If Msg 1 was successful, the probability that it may not receive Msg 2 is \((1-e)^{-i}\), where ‘\(i\)’ is the number of times the device re-transmits Msg 1 preamble \cite{3GPP37868}. If Msg 2 was successfully received, the probability of successful delivery for Msg 3 and Msg 4 is assumed as 90\% with non-adaptive HARQ with a maximum of 5 re-transmissions. The total transmission delay \(T_{total}\) is given as \(T_{total} = T_{wait} + T_{Msg \hspace{1mm}1} + T_{Msg\hspace{1mm}2} + T_{Msg\hspace{1mm}3} + T_{Msg\hspace{1mm}4}\), where \(T_{wait}\) is the Msg 1 wait time for successful preamble transmission and the rest are given in Table I. If the device fails the RACH procedure or does not receive any response to Msg 1, it will re-attempt after power ramping to \(P_{Tx}\) = min\{\(P_{max}\), (PL+\(P_{i}\)+ (C-1) \(\times\) step)\}, where \(P_{max}\) = 14 dBm is device transmit power, \(P_{i}\) = -104dBm is the initial received target power, step = 2 is the power ramping step size, C is the number of RACH attempts.

\begin{table}[h]
\vspace{-0.95em}
\caption{simulation parameters.}
\begin{tabular}{ll}
\toprule
\textbf{Parameter} & \textbf{Value}\\ \hline
Frequency & 2.6 GHz     \\
Channel bandwidth (BW)        & 5 MHz\\
Number of MTC devices & 5K\\ 
Arrival distribution (uRLLC) & Beta  (T=10 sec)\\ 
Arrival distribution (non-uRLLC) & Uniform  (T=30 sec) \\ 
PRACH configuration index & 6\\ 
Total number of preambles (\(N_{pre}\)) & 54\\ 
Maximum preamble transmissions (\(Max_{pre}\)) & 10\\ 
Number of UL grants per RAR & 3\\ 
Number of CCE allocated per PDCCH & 16\\ 
Number of CCE per PDCCH & 4\\ 
RA response window size \((RAR_{window})\) & 5 ms\\ 
mac-contention resolution timer & 48 ms\\ 
Back-off-indicator (BI) & Uniform (0,20) ms\\ 
HARQ probability for Msg 3 and Msg 4& 10\%\\ 
Max HARQ for Msg 3 and Msg 4 & 5\\ 
Msg 1 transmission time \((T_{Msg\hspace{1mm}1})\) & 1 ms\\ 
Msg 2 transmission time \((T_{Msg\hspace{1mm}2})\) & 3 ms\\ 
Msg 3 transmission time \((T_{Msg\hspace{1mm}3})\) & 5 ms\\ 
Msg 4 transmission time \((T_{Msg\hspace{1mm}4})\) & 5 ms\\ \hline
\end{tabular}
\label{parameters}
\end{table}

The key performance indicators (KPIs) of interest in RACH performance evaluation are collision probability, average access delay and resource (preamble) utilization. The collision probability is defined as the ratio between the number of occurrences when two or more devices send a RA attempt using the same preamble and the overall number of opportunities (with or without access attempts) in the period. The average access delay can be evaluated through the CDF of the delay for each RA procedure between the first RA attempt and the completion of the RA procedure, for the successfully accessed devices. The preamble utilization is the ratio between the total number of used preambles and the overall number of opportunities (with or without access attempts) in the period. Finally, Table II validates the simulation results using our system model against 3GPP technical report \cite{3GPP37868} with minor differences. 

\begin{table}[h]
\caption{Simulation results (proposed simulator vs 3GPP \cite{3GPP37868}).}
\label{table_example}
\begin{tabular}{@{}lllll@{}}
\toprule
\textbf{Num of devices} & \textbf{5K}& \textbf{5K}& \textbf{10K}& \textbf{10K}\\
\textbf{KPI} & \textbf{3GPP}& \textbf{proposed}& \textbf{3GPP}& \textbf{proposed}\\   \hline
Collision probability(\%) & 0.45 & 0.48 & 1.98 & 1.95 \\ \hline
Avg. preambles & 1.43 & 1.4 & 1.45 & 1.42 \\ \hline
Avg. access delay (ms) & 29.06 & 28.98 & 34.65 & 33.62 \\ \hline   
\end{tabular}
\end{table}

%%%%%%%%%%%%%%%%%%%%%%%%%%%%%%%%%%%%%%%%%%%%%%%%%%%%%%%%%%%%%%%%%%%%%%%%%%

 \section{Performance Results}
We evaluate the performance in two distinct scenarios: uRLLC traffic only and mixed traffic with co-existence of uRLLC and non-uRLLC traffic. 
%Two cases are considered namely: uRLLC traffic only and mixed traffic with non-uRLLC and uRLLC traffic.   

\subsection {uRLLC Traffic Only} 
\begin{figure}[!htbp]
   \centering
        \includegraphics[scale=0.45]{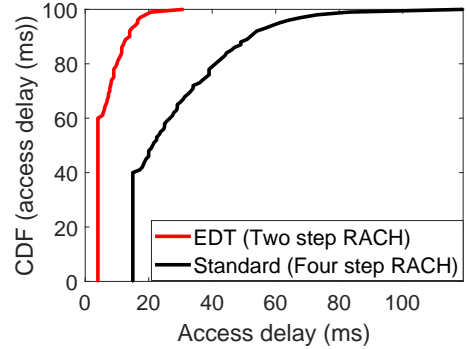}
      \caption{Performance evaluation of EDT versus standard LTE RACH.}
      \label{EDT_result}
\end{figure}

Considering a RA subframe occurs every 5 ms, Fig.  \ref{EDT_result} shows the access delay using EDT as compared to standard LTE. It can be seen that EDT significantly lowers the access delay from 29 ms to 6 ms at 50\% CDF. However, the Msg 1 collision probability is not effected. 

Table \ref{PP} shows the results using the proposed parallel preamble transmissions enhancement. If the device comes under the pico/femto cell coverage, it can perform RACH procedure on both the gNBs independently and simultaneously. The case of 0 femto cells reflects typical LTE scenario without any parallel preamble transmissions. Results from Table \ref{PP} show that collision probability improves by 50\% when the number of femto cells are 10.

\begin{table}[h]
\centering
\caption{Collision probability using parallel preambles.}
\label{PP}
\begin{tabular}{@{}cc@{}}
\toprule
\textbf{Number of femto cells} & \textbf{Collision probability (\%)}\\
\hline
0 (standard LTE) & 0.48\\ \hline
5 & 0.42 \\ \hline
8 & 0.34 \\ \hline
10 & 0.26 \\ \hline
12 & 0.22 \\ \hline
\end{tabular}
\end{table}

\begin{table}[h]
%\centering
\caption{Proposed enhancements KPIs for uRLLC only traffic.}
\begin{tabular}{lccc}
\toprule
\textbf{KPI} & \textbf{LTE} & \textbf{EDT + PP }  & \textbf{EDT + PP + EBF} 
 \\
\hline
Collision probability (\%) & 0.48 & 0.04 & 0.01\\ \hline
Average preamble  & 1.43 & 1.2 & 1.09\\ 
transmissions & & & \\ \hline
Average access delay (ms) & 29.06 & 5.8 & 4.47 \\ \hline
\end{tabular}
\label{EDT+PP}
\end{table}

\begin{figure}[!htbp]
   \centering
        \includegraphics[scale=0.4]{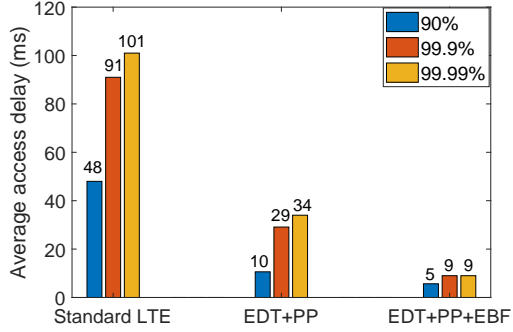}
      \caption{Access delay with proposed enhancements for uRLLC only traffic.}
      \label{uRLLC_final}
\end{figure}

When EDT is combined with parallel preamble (PP), which is referred to as EDT + PP in Table \ref{EDT+PP} and Fig.  \ref{uRLLC_final}, results show that the mean access delay and collision probability are significantly reduced as compared to the standard LTE. However, from Fig.  \ref{uRLLC_final}, the access delay for 99.99\% of the devices is still 34 ms which does not satisfy the 3GPP control plane target access delay of less than 10 ms. Thus, proposed enhanced back-off (EBF) scheme is further used, referred as EDT + PP + EBF in Table \ref{EDT+PP} and Fig.  \ref{uRLLC_final}, and results show that an access delay of 9 ms with a reliability of 99.99\% can be obtained in Fig.  \ref{uRLLC_final}. This is because EBF algorithm gives priority to the failed devices and lowers the back-off time. 

Table \ref{5G} shows the results for the 5G flexible numerology i.e. scaling of subcarrier spacing and mini-slots (where sym refers to symbols) as compared to standard LTE. Results show that the 3GPP target access delay can be achieved (i.e., less than 10 ms) either by using 60 
kHz subcarrier spacing with LTE slot configuration or using the LTE 15kHz frequency spacing but with mini slot of 2 symbols/slot.

%\begin{table}[h]
%\centering
%\caption{5G flexible numerology performance comparison with LTE.}
%\label{5G}
%\begin{tabular}{@{}lllll@{}}
%\toprule
%\textbf{Mean access delay (ms)} & \textbf{slot = }& \textbf{mini slot = }& \textbf{mini slot = }\\ 
%\textbf{collision probability (\%)} & \textbf{7 symbols} &\textbf{4 symbols} &\textbf{2 symbols} \\ \hline
%15 kHz & 29/0.48 & 14.7/0.45 & 6.9/0.45 \\ \hline
%30 kHz & 12.5/0.46  & 6.8/0.5 & 3.3/0.44 \\ \hline
%60 kHz & 6/0.47 & 3.14/0.43 & 1.68/0.46 \\ \hline
%120 kHz & 2.9/0.43 & 1.66/0.49 & 0.83/0.5 \\ \hline
%\end{tabular}
%\end{table}

%----------New----------
\begin{table}[h]
\centering
\caption{5G flexible numerology performance comparison with LTE.} \label{5G} \begin{tabular}{@{}lllll@{}} \toprule  \textbf{Subcarrier}  & \textbf{KPI} & \textbf{Slot}& \textbf{Mini slot}& \textbf{Mini slot }\\  \textbf{spacing} & & \textbf{7 sym} &\textbf{4 sym} &\textbf{2 sym} \\ \hline
15 & Mean access delay(ms) & 29 & 14.7 & 6.9 \\
 (kHz)  & Collision probability(\%) & 0.48 & 0.45 & 0.45\\ \hline
30 & Mean access delay(ms) & 12.5 & 6.8 & 3.3 \\
 (kHz) & Collision probability(\%) & 0.46  &0.5 & 0.44 \\ \hline
60 & Mean access delay(ms) &  6 & 3.14& 1.68 \\
  (kHz) & Collision probability(\%)  &  0.47 & 0.43 & 0.46 \\ \hline
120 & Mean access delay(ms) & 2.9 & 1.66 & 0.83 \\
  (kHz) & Collision probability(\%) & 0.43 & 0.49 &0.5 \\ \hline \end{tabular} \end{table}

In summary, results show that EDT combined with parallel preambles and enhanced back-off can meet the 3GPP uRLLC target requirements and is proposed in this paper. Also, 5G flexible numerology can be used.

\subsection {Mixed Traffic } 
A mixed traffic case with 5\% uRLLC and 95\% non-uRLLC devices is considered. In this case, priority needs to be maintained for the uRLLC devices to satisfy the latency and reliability requirements. Table \ref{reserved} shows that by reserving \(r\) = 3 preambles for uRLLC devices, provides guaranteed access with lower collision probability but with a lower reserved preamble utilization of only 38\%. However, on the extreme end if \(r\) = 1, the uRLLC devices contend for the single reserved preamble and the collision probability increases to 33\% with higher preamble utilization of 83\%. Thus, results show that fixing the number of reserved preambles can waste the valuable preamble resources.

\begin{table}[h]
\centering
\caption{KPI comparison of reserved preambles for mixed traffic.}
\label{reserved}
\begin{tabular}{@{}lllll@{}}
\toprule
\textbf{Reserved preambles (\(r\))} & \textbf{1}  & \textbf{2} & \textbf{3}  & \textbf{4} \\ \hline
uRLLC Collision probability (\%) & 33 & 0.97 & 0 & 0 \\ \hline
uRLLC 	preamble utilization (\%) & 83 & 57 & 38 & 29 \\ \hline
non-uRLLC Collision probability (\%) & 0.1 & 0.07 & 0.03 & 0.06\\ \hline
non-uRLLC 	preamble utilization (\%) & 3.4 & 3.1 & 3.1 & 3.1 \\ \hline
\end{tabular}
\end{table}

Results from Table \ref{DRP} show that EDT combined with the proposed algorithms of dynamic reserved preambles (DRP) and enhanced back-off (EBF) referred as EDT + DRP + EBF can reduce the mean access delay by 82\% (26 ms down to 4.5 ms) as compared to standard LTE. Table \ref{DRP} also shows that using reserved preamble referred as RP increases the collision probability for non-uRLLC type devices (from 0.11 to 1.06) which is not the case for the proposed EDT + DRP + EBF. The greatest benefit also comes due to the increase in the reserved preambles usage for uRLLC devices (from 23\% using RP to 57\% using EDT + DRP + EBF).

\begin{table}[h]
\centering
\caption{Proposed enhancements KPIs for mixed traffic.}
\label{DRP}
\begin{tabular}{@{}lllll@{}}
\toprule
\textbf{KPI   }         &     \textbf{ device type }         &\textbf{LTE} & \textbf{RP} & \textbf{EDT + DRP} \\ 
               &                           &             &             & \textbf{+ EBF} \\ \hline
Mean access&   uRLLC and     &      &             &      \\ 
delay (ms) & non-uRLLC    & 26.07        &  25              &    4.5   \\ \hline
Collision    &  uRLLC        & 0.05        & 0            & 0\\ 
\cline{2-5} 
probability (\%) &non-uRLLC  & 0.11      & 1.06         & 0    \\ \hline
Preamble      &  uRLLC       &             &   23          & 57 \\ 
\cline{2-5} 
utilization (\%)  &non-uRLLC &  2.10           &    3         &   7.6        \\ \hline
\end{tabular}
\end{table}

\begin{figure}[!htbp]
   \centering
        \includegraphics[scale=0.35]{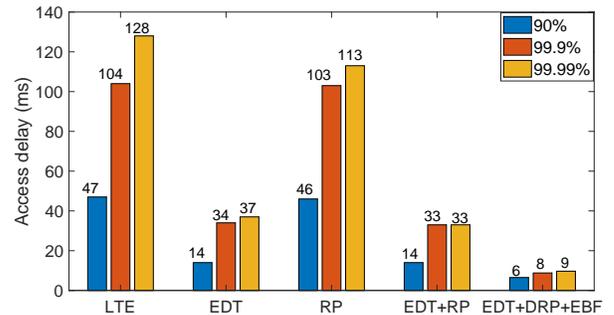}
      \caption{uRLLC access delay with proposed enhancements for mixed traffic.}
      \label{mixed_uRLLC}
   \end{figure}

\begin{figure}[!htbp]
   \centering
        \includegraphics[scale=0.35]{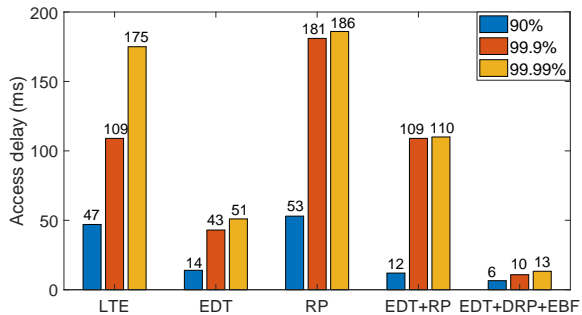}
      \caption{Overall access delay with proposed enhancements for mixed traffic.}
      \label{mixed_all}
   \end{figure}

The CDF of the access delay for proposed enhancements is shown in Fig.  \ref{mixed_uRLLC} for uRLLC devices and Fig.  \ref{mixed_all} for both uRLLC and non-uRLLC devices. As shown, the overall access delay for 99.99\% of the uRLLC devices is 9 ms (Fig.  \ref{mixed_uRLLC}) using EDT + DRP + EBF which satisfies the 3GPP target of 10 ms. Also, the overall access delay for 99.99\% of all traffic is reduced to 13 ms (Fig.  \ref{mixed_all}) using proposed EDT + DRP + EBF as compared to 175 ms using standard LTE. This indicates that the maximum Msg 1 attempts for the uRLLC devices is 1-2 to be successful. 
   
In summary, for a mixed traffic case, EDT  combined with dynamic reserved preambles and enhanced back-off, as proposed in this paper, can  meet  the 3GPP  uRLLC  target requirements. 
%%%%%%%%%%%%%%%%%%%%%%%%%%%%%%%%%%%%%%%%%%%%%%%

\section{Concluding Remarks}
This paper evaluated the performance of RA enhancements for 5G uRLLC. Moreover, it proposed new  RA enhancements for 5G uRLLC from an FoF perspective. These techniques have been specifically designed to meet the 3GPP stringent access delay and reliability requirements for uRLLC traffic in the control plane. The adoption of parallel preamble transmissions, dynamic reserved preambles and enhanced back-off ensure preamble success in the first attempt and complete the RACH procedure early. These techniques when combined with EDT and 5G flexible numerology reduce the access delay by 90\% as compared to the standard LTE solution. Performance evaluation further demonstrate that these enhancements outperform the existing enhancements in terms of collision probability, preamble utilization and access delay. Besides, dynamic reserved preamble with enhanced back-off has shown to reduce the access delay even for non-uRLLC devices which is particularly attractive in FoF environments. 

%%%%%%%%%%%%%%%%%%%%%%%%%%%%%%%%%%%%%%%%%%%%%%%%%%%%%
\section{Acknowledgement}
The work presented in this paper is partly funded by the EU's Horizon 2020 research and innovation programme under grant agreement No 761745 and the Government of Taiwan. 
%The authors acknowledge the contributions of all Clear5G partners in discussions leading to the drafting of this paper. 


%\section*{Acknowledgment}

%This work has been partially supported by the ICT-ACROPOLIS Network of Excellence, FP7 project %no. 257626, www.ict-acropolis.eu.



\bibliographystyle{IEEEtran}

\bibliography{IEEEabrv,sample}
%\begin{thebibliography}{1}
%\end{thebibliography}

\end{document}